\documentclass[prl,twocolumn]{revtex4}
\usepackage{graphicx}
\renewcommand{\narrowtext}{\begin{multicols}{2} \global\columnwidth20.5pc}

\renewcommand{\v}[1]{{\bf #1}}

\newcommand{\w}{{\omega}}

\def\be{\begin{eqnarray}}
\def\ee{\end{eqnarray}}

\newcommand{\Eq}[1]{Eq.~(\ref{#1})}

\newcommand{\ra}{\rightarrow}

\begin{document}
\draft

\title{On the Charge Ordering Observed by Recent STM Experiments}
\author{Henry C. Fu$^{a}$, J.C. Davis$^{b}$ and Dung-Hai Lee$^{a,c}$}
\affiliation{${(a)}$Department of Physics,University of California
at Berkeley, Berkeley, CA 94720, USA}\affiliation{${(b)}$ LASSP,
Department of Physics, Cornell University, Ithaca, NY 14850 USA.}
\affiliation{${(c)}$ Material Science Division, Lawrence Berkeley
National Laboratory,Berkeley, CA 94720, USA.}

\date{\today}
\begin{abstract}
We present a perspective on the recent discoveries
of possible charge ordering in underdoped
$Bi_2Sr_2CaCu_2O_{8+\delta}$ and $Na_xCa_{2-x}CuO_2Cl_2$ by
high-resolution scanning tunnelling spectroscopy.
\end{abstract}

\maketitle
\parindent 10pt

 In an STM study of optimally doped
$Bi_2Sr_2CaCu_2O_{8+\delta}$ ($Bi2212$) at finite applied magnetic
field, Hoffman {\it et al} discovered that in the vicinity of
vortex cores the STM tunnelling conductance exhibits a
checkerboard spatial dependence\cite{jenny}, with Fourier peaks at
$\v Q=(\pm 2\pi/\lambda,0), (0, \pm 2\pi/\lambda)$ and
$\lambda\approx 4.3 a$ ($a$ is the lattice constant of the copper
oxide plane). Subsequently Howald {\it et al} studied optimally
doped $Bi2212$ below $T_c$ at zero field, and reported observation
of a nondispersive, directionally asymmetric tunneling conductance with Fourier peaks at $\v Q=(\pm 2\pi/\lambda,0), (0, \pm 2\pi/\lambda)$ and $\lambda\approx 4a$ for low energies ($|V| < 25meV$)\cite{howald}. However this result was not confirmed by
later experiments\cite{hoffman,kyle,ali}. In
Ref.\cite{hoffman,kyle,ali} high resolution STM studies of
near-optimally doped $Bi2212$ below $T_c$ revealed rich
bias-dependent quasiperiodic modulations of the STM tunnelling
conductance.  This phenomenon was explained as the quantum
interference of quasiparticle de Broglie
waves\cite{hoffman,kyle,wang}. Ref.\cite{stevereview} emphasized
the deviation in the dispersion of the Fourier peaks in the
$(1,0)$ and $(0,1)$ crystallographic directions from the simplest
quasiparticle interference predictions.

In Ref.\cite{ali} Vershinin {\it et al} have also reported results
for $T> T_c$. Interestingly, at low energies they observed
bias-independent conductance modulations with Fourier peaks at $\v
Q=(\pm 2\pi/\lambda,0), (0, \pm 2\pi/\lambda)$ and $\lambda\approx
4.7 a$\cite{ali}.

Two very recent low temperature STM experiments on underdoped
$Bi2212$\cite{stm1} and $Na_xCa_{2-x}CuO_2Cl_2$
($NaCCOC$)\cite{stm2} reported  
bias independent quasiperiodic modulations of the local
conductance over a wide range of energy. These modulations have
Fourier peaks at $\v Q=(\pm 2\pi/\lambda,0), (0, \pm
2\pi/\lambda)$, with $\lambda\approx 4.5 a$ for $Bi2212$, and
$\lambda\approx 4 a$ for $NaCCOC$. Inspection of the real space
modulation patterns shows that they are {\it in phase} upon bias
reversal\cite{ali,stm1,stm2}. Ref.\cite{demler} and
Ref.\cite{zhang1} suggest that this bias symmetry contains vital
information about the origin of these phenomena. 
Here we would like to point out that this symmetry could appear
simply because the system is a doped Mott insulator.
Indeed, doping a Mott insulator quite commonly transfers spectral weight
(for both electrons and holes) proportional to the doping density from
above the charge gap to low energy. As a result, when the doping density
modulates in space, both the electron and hole spectral functions modulate
with the same phase.  

Despite the similarities between the phenomena reported in
Ref.\cite{stm1} and Ref.\cite{stm2}, the $Bi2212$ and $NaCCOC$
results are different in three important aspects. 1) The charge
period for $Bi2212$ is not an integral multiple of the underlying
lattice constant, while that of $NaCCOC$ is. 2) For $Bi2212$ the
$4.5 a$ modulation appears in $dI/dV$ spectrum only when $|V|\ge
65 mV$, while in $NaCCOC$ the signal appears right from zero bias.
3) For $|V|\le 30 mV$ $Bi2212$ exhibits bias-dependent
quasiparticle interference modulation\cite{hoffman}, while such a
signal is difficult to discern in $NaCCOC$. Due to these
differences it is possible that the phenomenon observed in
$NaCCOC$ is due to true charge order, while that observed in
$Bi2212$ is due to a yet-uncondensed charge order\cite{stm1,stm2}.

The above STM experiments constitute an important set of
observations concerning the electronic structure of the cuprates.
The purpose of this paper is to present our interpretation of the
results in Ref.\cite{jenny,ali,stm1,stm2}.

For $NaCCOC$, the $dI/dV$ modulation patterns consist of clearly visible $4\times 4$ unit
cells over a wide range of bias voltages
\cite{stm2} ($-150 mV\le V\le 150 mV$). A caricature of the real
space pattern seen in Ref.\cite{stm2} is shown in Fig.1a, where
bright/dark marks regions of high/low differential conductance. In
Ref.\cite{stm2} it is proposed that this spatial structure is due
to the formation of a Wigner crystal of doped holes. In the following we
corroborate this proposal with a theoretical analysis.

\begin{figure}
\includegraphics[width=8cm,height=3.25cm]{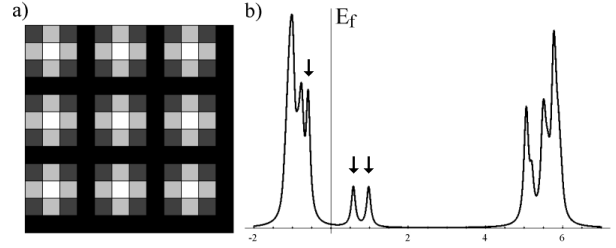}
\caption{a) Caricature of the observed spatial modulation in
$dI/dV$ in Ref.\cite{stm2}.  b) Spatially averaged
spectral function of the hole crystal. The spectral peaks
marked by arrows correspond to states whose wavefunction is
peaked at the solitons.}
\end{figure}

We start from a very underdoped situation where $x=1/16$. This is
the ideal doping density for a $4\times 4$ unit cell hole crystal\cite{hor}. 
We interpret the bright region in each unit cell as having a high
hole density (from a single hole delocalized
in that region), 
and the dark region as having a low hole density. In the
literature it has been proposed that doping a Mott insulator creates solitons (or spin
bags/composite quasiparticles)\cite{spinbag, bob, gan}.
We believe that the above bright and dark regions reveal the internal structure of each
soliton.
Recently Chen {\it et al} proposed that at commensurate doping
densities a Wigner crystal of {\it hole pairs} can form\cite{zhang}.
We believe that due to Coulomb repulsion such a state is energetically unstable.
Moreover, even with a strong short-ranged attractive interaction to overcome the
Coulomb interaction, we would expect segregation into
hole-rich and hole-poor regions rather than crystallization of hole pairs.

To demonstrate the possibility of a soliton crystal we perform a
variational study of the usual Hubbard model with Coulomb and nearest
neighbor exchange interactions added\cite{note}, \be
H&=& H_{Hubbard} + \sum_{(i,j)} V_{ij} n_i n_j+
J\sum_{<ij>}\v S_i\cdot\v S_j.\label{hubbard}\ee  The variational
ansatz we use is the most general Slater determinant which allows
spatially inhomogeneous charge density, spin density, bond-current
density, and superconducting pairing.  Technically this is
equivalent to factorizing the four-fermion terms in \Eq{hubbard}
into all possible quadratic combinations, and then solving the
resulting quadratic Hamiltonian self-consistently.
The following results are obtained numerically for an $8\times 8$ lattice with four holes and periodic boundary conditions.

It is quite encouraging that such a calculation does indeed yield
a crystalline arrangement of the doped holes for a reasonable
choice of parameters ($t=-0.3 eV, U= 6 eV, J=0.3 eV, V_c=0.3 eV$,
where $V_c$ is the repulsion energy two holes experience at the
nearest neighbor distance)\cite{note2}. We have checked that as
long as the Coulomb interaction exists, the stability of the
soliton crystal is not affected by moderate modification of the above parameters. In all the cases we have studied the
superconducting pairing and orbital current order are both absent
when the holes crystallize.  On the contrary we find that a
non-zero antiferromagnetic order coexists with the hole crystal.

In Fig.1b we show the spatially averaged spectral function.  The spectral peaks marked by arrows
correspond to states whose wavefunctions are peaked at the soliton positions shown by the hole
density (Fig.2a). The norm-squared wavefunctions of these states is reflected in the spatial
variation of the local spectral function within $\pm 150 meV$ of the Fermi level, as shown in
Fig.2b. Although the detailed distribution of the spectral weight within the unit cell differs
for the positive and negative bias, the gross structure (i.e. the fact that the overall spectral
weight peaks near the soliton) is the same. In that regard the spatial structure is similar to
that observed in Ref.\cite{stm2}. Contrary to experiment, in our calculation the dark region
basically has the LDOS spectrum of the undoped insulator. We attribute the $~150 meV$ peak
in the dark region seen in Ref.\cite{stm2} to the extra holes running on top of the
Wigner crystal background which are absent in our calculation.

We also note that when a soliton is produced on a Neel background, its location (on the A or B sublattice of the antiferromagnetic order) depends on the spin of the removed electron. For example, if removing a spin up electron produces a soliton on the A sublattice, then removing a spin down electron will produce a soliton on the B sublattice. Moreover, our calculation shows that after removing an electron of a given spin, an electron of the opposite spin is always deeply bound to the soliton. This result is consistent with the notion that a soliton is a composite particle of a charged void and a
spin\cite{bob}.  

\begin{figure}
\includegraphics[width=7.5cm,height=2.8cm]{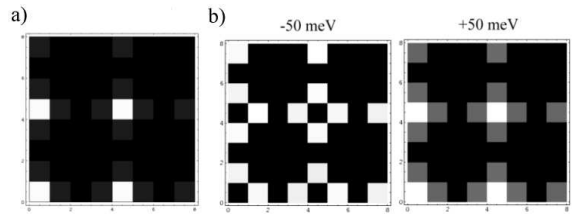}
\caption{a)Hole density distribution of the hole crystal. b)Spatial modulation of the local spectral function. The images shown here are typical for $-150 meV \le E\le 150 meV$. Although the detailed distribution of the spectral weight within the unit cell differs for the positive and negative bias, the gross structure (i.e. the fact that the overall spectral weight peaks near the soliton) is the same.}
\end{figure}

A weakness of the above variational study is that it underestimates
the quantum fluctuation of spins. To study the internal structure
of each soliton more carefully, we diagonalize the t-J model
on a $4\times 4$ plaquette with periodic boundary conditions.
The Hamiltonian we use for this study is

\begin{eqnarray}
 H=\sum_{<ij>}-t P_i(c^{\dagger}_{i\sigma} c_{j\sigma} +
 h.c.)P_j
    +J({\bf S_i}\cdot{\bf S_j}-{1\over 4}n_i n_j).
\label{tj}
\end{eqnarray}
In the following we choose $t=0.3 eV$, $J=0.1 eV$.
  Motivated by the experimental findings we restrict the hole to
a $3\times 3$ plaquette in the unit cell.
We compute the low-energy positive-bias\cite{why} electron spectral function
\be A(\v x,\omega)=\sum_{\alpha\sigma}\delta( \omega -E^{0h}_\alpha+
E_0^{1h}+\mu)|< \Psi^{0h}_\alpha | c^{\dagger}_{\v
x\sigma}|\Psi_0^{1h}
>|^2\label{spectral}\ee
Here $\mu$ is the chemical potential, $| \Psi_0^{1h}>$ is the
one-hole ground state with energy $E_0^{1h}$, and
$|\Psi^{0h}_\alpha
>$ are the eigenstates of the half-filled system, with energy
$E_{\alpha}$ at most $2J$ above the half-filled ground
state\cite{hybridization}. Fig.3 shows the hole density and $A(\v
x,\w)$ at the three inequivalent sites of the central $3\times 3$
plaquette. The hole density displays a similar spatial structure
to that seen in the Hubbard model calculation. The tunnelling
spectrum at the brightest site (i.e. site 3) exhibits two peaks.
To understand these two peaks we note that in a $4\times 4$ system
the one-hole ground state is a spin doublet. The unpaired spin 1/2
is meant to mimic the dangling spin in the soliton. After the
tunnelling process the spin from the added electron can form a
singlet or a triplet with the original soliton spin. The
lower(higher) peak corresponds to the lowest energy spin doublet
$\ra$ spin singlet(triplet) transitions.  The peaks seen in Fig.3
are similar to those observed in the bright region of the $4\times
4$ unit cell in Ref.\cite{stm2}. (Because we constrained the hole
to the central $3\times 3$ region, we cannot address the $~150
meV$ peak observed in the dark region of the $4\times 4$ unit
cell.)

\begin{figure}
\includegraphics[width=7.2cm,height=4.5cm]{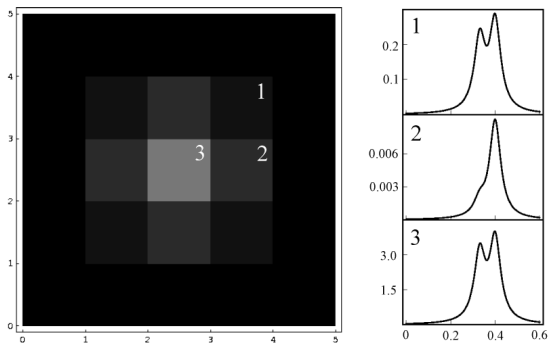}
\caption{Hole distribution and positive bias local spectral function. The center of
each plaquette is a $Cu$ lattice site. The seven boundary sites of the $4\times 4$
unit cell are replicated using periodic boundary conditions. The zero of the horizontal axis can be shifted by the chemical potential $\mu$ (here $\mu=0$).}
\end{figure}

In the above discussions we have concentrated on doping density
$x=1/16$, where the system is a charge insulator. The actual
doping density in the sample studied in Ref.\cite{stm2} is likely
to be higher\cite{uncertain}.  We assert that when $x$ deviates
from $1/16$ the extra holes are delocalized on top of the Wigner
crystal background. These mobile holes metallicize the system, and
at low enough temperatures such a metal can become a
superconductor. When that happens the Wigner crystal order
coexists with superconductivity. Clearly an important question is
the range of stability (in doping) of the Wigner crystal.
Currently we do not have an explicit answer to this question.
However, we emphasize that in answering this question the effects
of lattice pinning (including the effect of periodic potential as
well as elastic lattice relaxation) are extremely important. We
propose that the $4 \times 4$ unit cell Wigner crystal with extra mobile holes is adiabatically connected to the system with partially gapped Fermi surface near the antinodes discussed later in this paper\cite{zx}.


In the following we interpret the STM results for $Bi2212$
\cite{hoffman,stm1,jenny,ali}.  It is widely accepted that at
optimal doping and lowest energies the only important elementary
excitations are nodal quasiparticles. (The phase fluctuations are
gapped due to the 3D Coulomb interaction.) As the excitation
energy increases, antinodal quasiparticles and vortex-antivortex
excitations join the list. The smallest vortex-antivortex pair is
a roton, which appears as a (damped) pole of the dynamic
density-density correlation function. Because the signatures of
charge ordering at wavevectors $\v Q\approx (\pm\pi/\lambda,0),
(0,\pm\pi/\lambda)$ are observed in the underdoped regime, we
assert that there exist four discrete roton minima at wavevectors
$\v Q_{roton}\approx (\pm\pi/\lambda,0),
(0,\pm\pi/\lambda).$ 
These roton minima enhance the DC charge susceptibility at $\v
Q_{roton}$. 
We believe that this enhancement is due to the particle-hole
scattering across the nearly nested Fermi surface in the antinodal
region (with nesting wavevector $\v Q_{nesting}$, see the left panel of Fig.4), so that $\v Q_{roton} \approx \v Q_{nesting}$.

Due to this enhancement, Friedel-like modulations in the electron
density at $\v Q_{roton}$ are induced by disorder.  These
modulations can in turn scatter the quasiparticles and give rise
to enhanced quasiparticle interference modulations at wavevector
$\v Q_{roton}$. The lowest order such process is illustrated in Fig.4a\cite{wang,others,demler,stevereview}. 
We believe that the bias-independent local conductance modulations
in Ref.\cite{jenny,ali} of optimally doped $Bi2212$ are due
to the scattering described above.

In addition to Fig.4a the process
depicted in Fig.4b can also scatter quasiparticles with momentum
transfer $\v Q_{nesting}$. This process is resonantly enhanced if
the quasiparticle energy is greater than the excitation energy of
the roton\cite{subir}.
Given the fact that the $\lambda\approx 4.5 a$ modulation
phenomenon in Ref.\cite{stm1} appears for  $|V|>65 mV$, we think
it quite likely it is due to the scattering process shown in
Fig.4b. Finally, as pointed out by Zhou {\it et al}\cite{zhou},
the process shown in Fig.4c (without impurity involvement) can be
very effective in destroying the coherence of antinodal
quasiparticle excitations in ARPES studies\cite{arpes}.

\begin{figure}
\includegraphics[width=7.2cm,height=4.5cm]{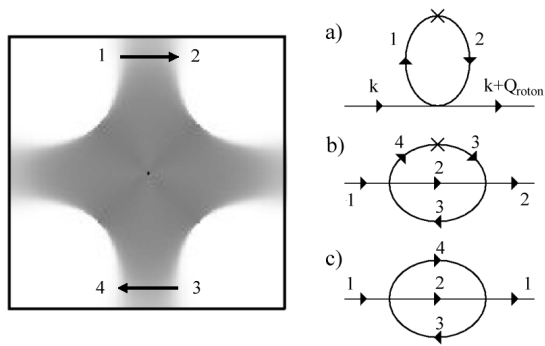}
\caption{The non-linear processes by which a quasiparticle is
scattered by the Friedel oscillation set up by an impurity.}
\end{figure}

As a function of decreasing $x$ the roton minima deepen. When
these minima touch zero, roton condensation occurs. This marks the
onset of charge order. In an isotropic superfluid such as $^4He$
the onset of crystallization coincides with the loss of
superfluidity. However, in a system like the $NaCCOC$  where the
charge order is commensurate with the underlying crystalline
lattice, interstitial/vacancy superconductivity can coexist with
Wigner crystal order due to lattice pinning effects.
In such a coexistence phase, the antinodal quasiparticles are
strongly scattered by $\v Q_{roton}$, hence a gap can open up in
the antinodal region. However since $\v Q_{roton}$ does not
connect the nodes the nodal quasiparticles can remain gapless if
the potential from charge order is not too strong. Under these
conditions the nodal quasiparticle peak can survive in the ARPES
spectra\cite{zxnaccoc}. Upon further decreasing $x$
superconductivity eventually goes away. The loss of
superconductivity is triggered by the Bose condensation of {\it
single} vortices.  With single vortices condensed all doped holes
localize and the system becomes an insulator.  After localization
of holes the antiferromagnetic order is presumably restored.
Because a real system has disorder, the Wigner crystal order
discussed above can only have finite-range correlations due to the
Imry-Ma physics.

Is the charge ordering discussed here and in
Ref.\cite{ali,stm1,stm2} responsible for the opening of the
pseudogap?  As has been pointed out in Ref.\cite{stm1} this
crucially depends the relation between the charge freezing
(crossover) temperature $T_{charge}$ and the pseudogap temperature
$T^*$.  If $T_{charge}=T^*$ a strong argument can be made that the
charge order actually drives the pseudogap formation. However, in our opinion it is likely that the pseudogap is caused by a different
mechanism (e.g. spin singlet formation), and $T_{charge} < T^*$.

Finally, we note that like stripes\cite{tranquada}, the hole
crystal discussed in this paper represents yet another form of
charge order. We do not have much to say about how the system
chooses one form or the other.  What is important is that the
cuprates seem to have a tendency to form some type of charge
order. Whether these charge orders have anything to do with
pairing is entirely unclear to us.

{\bf Acknowledgements}: JCD and DHL would like to thank T.Hanaguri,
J.E. Hoffman, E. Hudson, K. Lang, C. Lupien, K. McElroy,
and R.Simmonds for numerous discussions. We thank A. Yazdani for valuable discussions and sharing his unpublished data with us. We also thank A. Lanzara, G.-H. Gweon, S.A. Kivelson, P.A. Lee, A. Seidel, Z.-X. Shen, and
X.-J. Zhou for valuable discussions. DHL is supported by DOE grant
DE-AC03-76SF00098. HF thanks Z. Hussain of the Advanced Light
Source for support.

\widetext

\begin{references}
\bibitem{jenny} J.E. Hoffman, {\it et al}, Science {\bf 295},466 (2002).
\bibitem{howald} C. Howald {\it et al}, Phys. Rev. B {\bf67}, 014533 (2003).
\bibitem{hoffman}J. E. Hoffman, et al. Science 297, 1148 (2002).
\bibitem{kyle} K. McElroy, et al  Nature 422, 520 (2003).
\bibitem{ali} M. Vershinin {\it et al}, 10.1126/science.1093384 (Science Express
Reports), February 12, 2004.
\bibitem{wang} Q. Wang and D.-H. Lee, Phys. Rev. B 67,
020511 (2003).
\bibitem{stevereview} S.A. Kivelson {\it et al}, Rev. Mod. Phys. {\bf75} 1201 (2003).
\bibitem{stm1} K. McElroy {\it et al}, to be published.
\bibitem{stm2} T. Hanaguri {\it et al}, submitted to Nature.
\bibitem{demler} D. Podolsky, E. Demler, K. Damle and B.I. Halperin, Phys. Rev. B {\bf67} 94514 (2003).
\bibitem{zhang1} H.-D. Chen {\it et al}, cond-mat/0402323.
\bibitem{hor} F. Zhou {\it et al}, Supercond. Sci. Technol. {\bf16}, L7 (2003) proposes that cuprates may be charge ordered at commensurate doping densities.
\bibitem{spinbag} J.R. Schrieffer, X.-G. Wen, S.-C.Zhang, Phys. Rev. Lett. {\bf 60} 944 (1988).
\bibitem{bob}P Beran {\it et al} Nucl. Phys. B, {\bf473}, 707
(1996); R.B. Laughlin, Phys. Rev. Lett. {\bf79}, 1726 (1997).
\bibitem{gan} J. Gan, D-H Lee, P. Hedeg\aa{}rd, Phys. Rev. B {\bf54}, 7737 (1996).
\bibitem{zhang} H.-D. Chen {\it et al}, Phys. Rev. Let. {\bf89}, 137004(2002); H-D. Chen {\it et al},
cond-mat/0312660.
\bibitem{note} We add the exchange term because without it the variational study precludes superconducting pairing.
\bibitem{note2} Our choice of $J$ here is bigger than the widely accepted value $J\approx 0.1 eV$ in t-J model studies. This is because in the
Hubbard model double occupancy is allowed.  The extra charge fluctuation suppresses the effect of $J$.
\bibitem{why} We focus on the positive bias because experimentally
the local spectrum is basically featureless for negative bias.
\bibitem{hybridization}  To make the computation tractable, we first
diagonalize the central $3\times 3$ plaquette
and then hybridize its
lowest energy states with the remaining spin sector of the $4\times 4$ system.
In the one-hole(half-filled) sector we hybridize eigenstates of the $3\times 3$ plaquette
which have energies within $1.3 J$($2.1 J$) of the ground state.
For this reason, in \Eq{spectral} we only sum over eigenstates of the half-filled $4\times 4$ lattice with energy $E_{\alpha}^{0h} < E_0^{0h} + 2 J$, where $E_0^{0h}$ is the ground state energy of the half-filled $4\times 4$ lattice.  For $\omega + \mu < 0.6$, the spectral function is not affected by changing the number of hybridized $3\times3$ plaquette states.
\bibitem{uncertain} The hole density is not precisely known because the sample surface doping may deviate from that in
the bulk.
\bibitem{zx} The existence of metallic charge density wave systems
with partially gapped Fermi surface was emphasized to us by Z.-X.
Shen.
\bibitem{others} L. Capriotti, D. J. Scalapino and R. D. Sedgewick.
(available at http://xxx.lanl.gov/abs/cond-mat/0302563).
\bibitem{subir} A similar scattering mechanism has been pointed
out by A. Polkovnikov {\it et al}, Physica C, 19, {\bf388-389},
(2003).
\bibitem{zhou} X.-J. Zhou {\it et al} to appear in Phys. Rev. Lett.
\bibitem{arpes}  A. Damascelli, Z. Hussain, Z-X, Shen, Rev. Mod. Phys. 75, 473 (2003) and references therein.
\bibitem{zxnaccoc} Y. Kohsaka {\it et al}, J. Phys. Soc. Jpn, {\bf72} 1018 (2003); F.Ronning {\it et al}, Phys. Rev. B {\bf67}, 165101 (2003).
\bibitem{tranquada} J.M. Tranquada {\it et al} Nature {\bf 375}, 561
(1995); J. Zaanen, O. Gunnarson Phys. Rev. B. {\bf40}, 7391
(1989); V.J. Emery, S.A. Kivelson Physica C {\bf209}, 597 (1993).
\end{references}
\end{document}